**RESEARCH**  Open Access

# DLP: towards active defense against backdoor attacks with decoupled learning process

Zonghao Ying[1,2] and Bin Wu[1,2*]

**Abstract**

Deep learning models are well known to be susceptible to backdoor attack, where the attacker only needs to provide a tampered dataset on which the triggers are injected. Models trained on the dataset will passively implant the backdoor, and triggers on the input can mislead the models during testing. Our study shows that the model shows different learning behaviors in clean and poisoned subsets during training. Based on this observation, we propose a general training pipeline to defend against backdoor attacks actively. Benign models can be trained from the unreliable dataset by decoupling the learning process into three stages, i.e., supervised learning, active unlearning, and active semi-supervised fine-tuning. The effectiveness of our approach has been shown in numerous experiments across various backdoor attacks and datasets.

**Keywords** Deep learning, Backdoor attack, Active defense

## Introduction

In recent years, deep learning technology has been widely applied across a number of domains, dramatically improving the efficiency of tasks such as object recognition (Eitel et al. 2015; Wang et al. 2015), semantic segmentation (Lateef and Ruichek 2019; Garcia-Garcia et al. 2018), speech recognition (Deng et al. 2013; Zhang et al. 2018), and machine translation (Costa-jussà and Escolano 2016; Vaswani et al. 2018). At the same time, the security of deep learning has also received attention.

Backdoor attack (Gao et al. 2020) has recently been proposed as a new attack paradigm. The attacker (or a malicious third party provider) can launch backdoor attack by providing a dataset injected with triggers. When the user trains a model directly on the dataset, the model will passively be implanted with a backdoor. In the testing phase, the model will predict a sample as a specific class if the trigger is present. Otherwise, it behaves normally. Since deep learning models do not behave differently without the trigger, the backdoor attack is very stealthy, which poses a threat to the practical application of deep learning. Defending against backdoor attacks effectively is an urgent issue.

A large amount of data is required to train deep learning models. However, due to the opaque nature of the data processing process, users have to fully trust datasets provided by others or collected from the Internet. Users can only take passive defensive measures for mitigation if attackers inject triggers into the dataset beforehand. Even though data-based defenses (Chen et al. 2019; Tran et al. 2018; Zeng et al. 2021) can detect poisoned samples, they cannot eliminate the threat since the backdoor is implanted into the model. There are also model-based defenses that can detect (Fields et al. 2021; Huster and Ekwedike 2021; Sikka et al. 2020) and mitigate (Li et al. 2021; Yoshida and Fujino

*Correspondence:
Bin Wu
wubin@iie.ac.cn
[1] State Key Laboratory of Information Security, Institute of Information Engineering, Chinese Academy of Sciences, Beijing, China
[2] School of Cyber Security, University of Chinese Academy of Sciences, Beijing, China

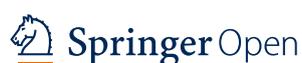

2020; Liu et al. 2018) backdoor threats, whereas the former can only discard the model after detecting the backdoor, while the latter requires further resources to repair. These passive defenses mitigate the threat of the backdoor attack, but they all share a common limitation: they cannot be implemented until the model has been trained. When a backdoor attack is detected and mitigated, significant training resources have already been wasted.

The paper proposes an active defense mechanism with a decoupled learning process (DLP) to mitigate this challenge. DLP decouples the standard learning process into three stages: supervised learning, active forgetting, and active semi-supervised fine-tuning. With decouple learning process, we can build a benign model from an untrustworthy dataset while balancing the effectiveness of backdoor removal and the usability of the model.

We argue that the model essentially learns both the backdoor task on the poisoned subset as well as the main task on the clean subset when it is trained on the tampered training set. We observe that in the early stages of standard learning, the model learns the backdoor task much better than it learns the main task. In response to this observation, we decoupled the standard learning process and set the first stage to supervised learning. Following this stage, clean and poisoned samples can be filtered using active learning (Ren et al. 2022). We then set up the active unlearning stage so that filtered samples are used as well as a gradient ascent algorithm in order to remove the backdoor. Furthermore, with the active semi-supervised fine-tuning phase, the usability of the model is further enhanced by combining the filtered clean samples with the semi-supervised approach. After three stages of learning, we can train a benign model on the tampered dataset. Since the decoupled learning process makes no assumption about the attack strategies for crafting poisoned samples, our proposed DLP is generic and applies to various attack methods.

Our main contribution are summarized as follows:

- We reveal that model will show significant differences in learning behavior by treating the process of training the backdoor model on the tampered dataset as the joint learning process of the main task and the backdoor task. In light of this observation, we developed an active learning-based strategy for filtering the two types of samples accurately.
- We propose a new defense against backdoor attacks called DLP, which actively trains benign models from tampered datasets. With DLP, the standard learning process can be decoupled into three stages, and a better usability-effectiveness trade-off can be achieved. The DLP is a simple yet powerful backdoor defense approach.
- We evaluate DLP against five well-known backdoor attacks. As a result of extensive experiments, we can consistently show that DLP can deliver state-of-the-art defensive performance.

The organization of the paper is as follows. Section "Related works" introduces the related work, including backdoor attack, backdoor defense, active learning, and semi-supervised learning. Section "Threat model" describes the threat model. Section "Characteristic of backdoor learning" illustrates our observations on the learning process of backdoored models. Section "Method" elaborates on the proposed DLP and its crucial components. Section "Experiments" evaluates the proposed attack. Section "Conclusion" concludes this paper.

## Related works
### Backdoor attacks

Deep learning models are subject to backdoor attack, which is a novel attack paradigm. It is possible for an attacker to embed a backdoor into a model by tampering with the training dataset. As a result of training on the tampered dataset, the backdoor is automatically inserted into the model. We refer to the training process in this scenario as backdoor learning. During the test phase, backdoor in the model will be activated by triggers, leading the model to make wrong predictions.

The BadNets proposed by Gu et al. (2017) demonstrated the feasibility of the backdoor attack for the first time, and the subsequent work improved the attack based on the results. Liu et al. (2018) achieved lower poisoning rates by designing triggers that activate specific neurons more efficiently. Chen et al. (2017) proposed a blended injection method to improve the stealth of poisoned samples. Several other works aim to improve the stealth of triggers by choosing particular patterns as triggers, such as reflections (Liu et al. 2020), raindrops (Zhao et al. 2022), and adversarial perturbations (Zhang et al. 2021). Some works explore new ways of backdoor learning, such as training the trigger generator and backdoored model simultaneously to make the model automatically learn the trigger patterns (Cheng et al. 2021; Nguyen and Tran 2020; Salem et al. 2022). Poison-label attacks and clean-label attacks are further divided based on whether the labels of the poisoned samples are modified. The latter does not require modifying the labels of the poisoned samples, making them harder to detect.

### Backdoor defenses

Many approaches have been proposed to defend against backdoor attacks. Due to the importance of data and model in deep learning, backdoor defense schemes can be divided into data-based defenses and model-based defenses. In data-based defenses, the defender can detect anomalous samples by neuron activations (Chen et al. 2019), spectral signatures (Tran et al. 2018), frequency features (Zeng et al. 2021), or training losses (Huang et al. 2022). There are also methods of reconstructing samples (Kwon 2021; Doan Bao et al. 2020) to undermine the validity of triggers in order to prevent backdoors from being activated. Model-based defenses employ discriminative features to detect backdoors, including signatures of model weights(Fields et al. 2021), transferability of adversarial perturbations (Huster and Ekwedike 2021), and counterfactual attribution (Sikka et al. 2020). Additionally, defenders can repair backdoored models through knowledge distillation (Li et al. 2021; Yoshida and Fujino 2020), pruning (Liu et al. 2018), or fine-tuning (Mu et al. 2022).

Li et al. (2021) first investigated how to train a benign model on the tampered dataset. After filtering the poisoned samples using training losses, they first improve the model performance by supervised learning and then remove the backdoor using filtered samples. Although they have developed a promising approach, it is not without limitations: firstly, their filtering method is insufficiently accurate, and secondly, the model's performance on the main task is further reduced when the backdoor is removed at the end of the process. Inspired by Li et al. (2021), we propose DLP, which can enhance defensive effects and compensate for the above weaknesses. We discuss the design details in Sect. "Methods".

### Active learning

The critical problem to be solved in the domain of active learning is how to maximize the model's performance by labeling a minimum number of samples. A crucial component of active learning is designing query strategies to identify samples that are difficult to predict and then handing them over to experts for labeling. There are two main principles behind classical query strategies: uncertainty and diversity. The uncertainty principle aims to identify samples for which the current model is least capable of predicting (Balcan et al. 2007; Holub et al. 2008). The diversity principle aims at finding samples that differ so that the information provided by the queried samples is comprehensive (Yang et al. 2015; Brinker 2003).

One of the main challenges in DLP is filtering out poisoned samples and clean samples from the training set. We argue that there are similarities between this and active learning because both need to design query strategies to find out the needed samples. It is important to note that the samples in our scenario are labeled, but due to the backdoor attacks, they are viewed as unlabeled samples. As a result, we need to design query strategies to filter some of them out and trust their label information. We argue that the most predictable samples are poisoned samples, and the most challenging samples are clean samples, so it makes sense to introduce active learning to solve the challenge.

### Semi-supervised learning

Large amounts of high-quality labeled samples are necessary to train deep learning models well. Labeled samples, however, are often difficult and expensive to obtain in many real-world scenarios. In contrast, unlabeled samples are usually readily available and large. Semi-supervised learning aims to improve a model's performance by using many unlabeled data and a small number of labeled samples. There are several semi-supervised learning methods, including pseudo-labeling methods (Lee 2013; Blum and Mitchell 1998), consistency regularization methods (Sajjadi et al. 2016; Miyato et al. 2019), graph-based methods (Iscen et al. 2019; Chen et al. 2020), and hybrid methods (Berthelot et al. 2019; Sohn et al. 2020).

In this paper, we introduce semi-supervised learning to address another challenge with DLP, namely how to improve further model performance given only a tiny number of high-confidence labeled samples.

### Threat model

We follow the standard thread model of backdoor attack, in which the attacker controls the dataset, and the user trains the model on unreliable dataset (specifically provided from the attacker). The model will exhibit targeted misclassification following training when presented with samples containing the trigger.

### Attacker's capacity and goals

The attacker's goal is to provide a tampered dataset, where the model trained from will be passively implanted backdoor. In this work, we consider the attacker has the maximum capability. With full access to the training dataset, the attacker can inject triggers into the dataset using any state-of-the-art backdoor attack strategy.

### Defender's capacity and goals

It is assumed that the defender can only control the model's training process and does not know the attack pattern, target type, or other details of the backdoor attack. The assumption is consistent with a typical deep learning training scenario.

The defender aims to train benign models from the tampered dataset. For a suitable defense mechanism to accomplish this goal, the following properties are necessary:

- Comprehensive: the defense can cover various backdoor attacks, regardless of trigger designs, injection methods, poisoning rate, and other factors that might be involved.
- Usability-preserving: the defense has a negligible impact on the model performance of the main task. Particularly if the dataset is not tampered, the model trained with defense method should perform similarly to the model trained using standard method.

### Formulation

This paper focuses on a typical class of backdoor attacks, namely single-target attacks (also known as all-to-one attacks), which are more efficient. Any input with a trigger is recognized by the model as the same label in a single-target attack, i.e., the target label is unique. Given original training dataset $D_o = \{x_i, y_i\}_{i=1}^n$, poisoning rate $\alpha$, and target class $y_t$, the original training dataset $D_o$ can be divided into target subset $D_t = \{x_j, y_j\}_{j=1}^m$ and clean subset $D_c = \{x_q, y_q\}_{q=m+1}^n$, where $m = n \times \alpha$. The attacker will inject triggers into all samples from $D_t$ according to specified attack strategy, and obtained poisoned subset $D_p = \{x_j', y_t\}_{j=1}^m$, where $x_j' = x_j \otimes \Delta$, $\otimes$ refers to injection method and $\Delta$ refers to the trigger. We call this process dataset poisoning. The attacker then releases the tampered dataset $D_m (D_m = D_p \cup D_c)$.

Once the user has trained the model with the dataset $D_m$, the model will be implanted with a backdoor. In the test phase, the model will behave normally on clean samples while predicting poisoned samples as target class:

$$\begin{cases} y_i = f_w(x_i), \\ y_t = f_w(x_i'), \end{cases} \quad (1)$$

where $f_w$ is backdoored model, $w$ are model parameters, $y_i$ is ground truth label of $x_i$, $x_i'$ is the sample injected with the trigger.

That is the whole process of the backdoor attack. The defender aims to train a benign model $f_{w^*}$ from $D_m$. The benign models give consistent predictions regardless of the sample's presence or absence of triggers:

$$y_i = f_{w^*}(x_i) = f_{w^*}(x_i') \quad (2)$$

where $w^*$ refers to the parameters of the benign model. Our strong threat model ensures the practical usage of DLP in real-world settings.

### Characteristic of backdoor learning

The standard learning procedure of model on the tampered dataset $D_m$ as follows:

$$\min_w \frac{1}{n} \sum_{(x,y) \in D_m} L_1(f_w(x), y) \quad (3)$$

where $L_1$ is supervised learning loss function(e.g. cross-entropy loss function). The optimization of the Eq. 3 can be realized by backpropagation with stochastic gradient descent. The training process on $D_m$ can be further refined as the training of the model on the clean subset $D_c$ and the training on the poisoned subset $D_p$:

$$\min_w \left( \frac{1}{m} \sum_{(x,y) \in D_p} L_1(f_w(x), y) + \frac{1}{n-m} \sum_{(x,y) \in D_c} L_1(f_w(x), y) \right) \quad (4)$$

We train the WideResNet-16-1 (Zagoruyko and Komodakis 2016) on the CIFAR10 (Krizhevsky and Hinton 2009) with representative backdoor attacks, BadNets (Gu et al. 2017) and SIG (Barni et al. 2019) respectively. These two attacks are typical strategies of poison-label attack and clean-label attack. We set the poisoning rate $\alpha$ to 10% and the batch size to 128.

Figure 1 shows the changes in training accuracy for poisoned and clean samples during backdoor learning. In the early stage of training, poisoned samples demonstrate greater accuracy than clean samples. This phenomenon suggests that the model learns the backdoor task faster than the main task. To master the backdoor task, the model only needs to learn the mapping of triggers to target classes. In order to enhance the effectiveness of backdoor attacks, attackers tend to design triggers into easily learnable patterns, such as a fixed simple image (Gu et al. 2017) or an optimized set of pixels (Liu et al. 2018). As a result, the backdoor task can be learned faster when both tasks are learned simultaneously. Our observations are supported by Arpit et al. (2017).

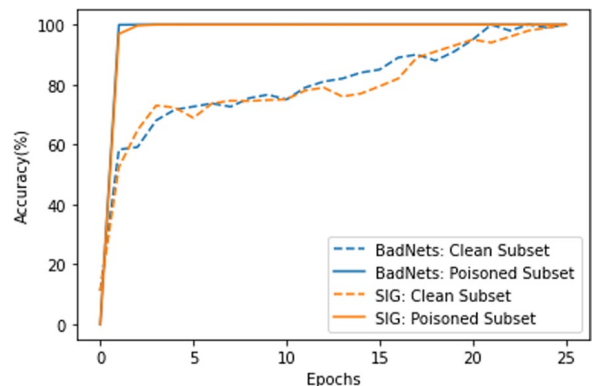

**Fig. 1** Training accuracy of clean subset and poisoned subset

A model's prediction of samples will reflect the differences in learning behaviors described above. The model can predict poisoned samples more confidently, while clean samples have a lower confidence level. The active learning strategy we developed in DLP exploited this difference to filter out both types of samples.

## Method

Our proposed DLP is described in this section. We will start with the overview of DLP and its pipeline, followed by its critical component.

### Overview

We make the following improvements to compensate for the limitations of Li et al. (2021). Firstly, we introduce active learning and develop a method based on predictive entropy to filter out desired samples with higher confidence. Secondly, we filter out clean samples and poisoned samples separately for subsequent use. Thirdly, we adjust the order of backdoor removal and model fine-tuning. We first remove the backdoor using a filtered poisoned subset and then fine-tune the model using a filtered clean subset.

It is important to note that the DLP uses semi-supervised fine-tuning instead of supervised fine-tuning. Due to the unknown details of the attack, such as the poisoning rate, we cannot completely filter out poisoned samples. In the case of supervised fine-tuning, a backdoor will once again be implanted in the model. By contrast, we can obtain benign samples with high accuracy with DLP, allowing us to improve model performance by semi-supervised fine-tuning without introducing a backdoor. With the above strategy design, DLP can achieve the best tradeoff between attack success rate and clean accuracy.

Specifically, DLP involves decoupling the model's learning process into three stages. The first stage is supervised learning (SL), achieved by performing initial standard training on the whole tampered dataset. At the end of this phase, the model will overlearn the backdoor task but not fully learn the main task. With this differential behavior, DLP can filter out poisoned samples and clean samples with a high level of confidence. The second stage is active unlearning (AU), which finally achieves backdoor removal by maximizing the same loss function as last stage. Active semi-supervised fine-tuning (ASSFT) is the final stage, in which the filtered clean subset is viewed as a labeled dataset to improve model performance on the main task.

Figure 2 illustrates this pipeline. Section "Entropy-based filtering method" details the method of filtering samples, and Sect. "Active unlearning" describes our active unlearning method for backdoor elimination. Section "Active semi-supervised fine-tuning" discusses the active semi-supervised fine-tuning for improving the model's performance on the main task.

### Entropy-based filtering method

As discussed in Sect. "Characteristic of Backdoor learning", the model obtained from initial supervised learning has difficulty predicting clean samples while easily predicting poisoned samples. We argue that in this scenario, the problem of accurately filtering the two types of samples is similar to what active learning works to solve: finding the samples that are most difficult (and easiest) to predict for the model.

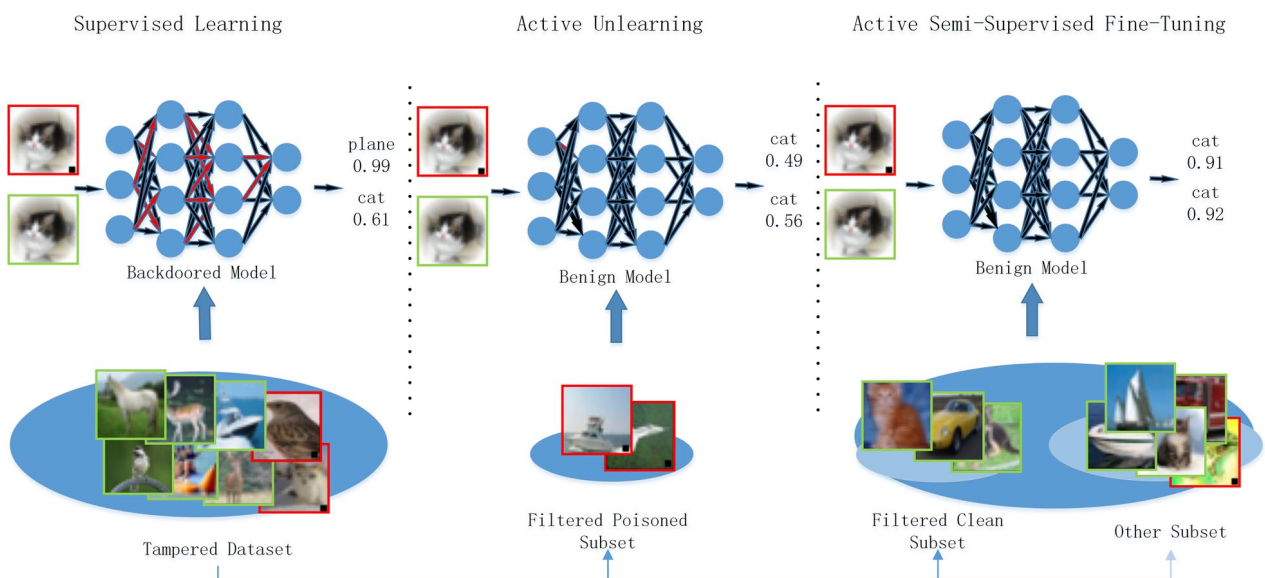

**Fig. 2** The main pipeline of DLP

We use Shannon entropy to represent prediction difficulty. The intuition behind this approach is that the model does not learn main tasks well, so predictions for clean samples are uncertain. Therefore, the probability of belonging to each class is almost the same in the corresponding prediction results, resulting in a higher entropy value. In contrast, poisoned samples have a lower entropy. The entropy of sample $x$ can be expressed:

$$H(x) = -\sum_{i=0}^{i=C} y^i \times log_2 y^i \quad (5)$$

where $y^i$ is the probability of sample $x$ belonging to class $i$, and $C$ is the total number of classes.

We apply Eq. 5 to all samples to calculate entropy and then sort them in ascending order. We filter samples according to the given filtering rate $\gamma$ based on the sorting results. The first $n \times \gamma$ samples are filtered out as the filtered poisoned subset $D'_p$. And the last $n \times \gamma$ samples are filtered out from each class as the filtered clean subset $D'_c$. Using this approach is both compatible with active learning's diversity principle and with semi-supervised learning's requirements.

### Active unlearning

The model can learn the critical features required to perform the corresponding task by minimizing a predefined loss function through supervised learning. In the scenario of backdoor attack, the model learns the features that are required for the backdoor task (called backdoor features) and those that are needed for the main task (called clean features) by minimizing the loss function under the poisoned subset and clean subset, respectively.

Removing the backdoor from the model is equivalent to having the model unlearn the backdoor features. Learning and unlearning are mutually antagonistic processes, so we can unlearn the backdoor features by maximizing the loss function on $D'_p$. Here is the optimization objective for this stage:

$$\max_{w'} \frac{1}{m} \sum_{(x,y) \in D'_p} L_1(f_{w'}(x), y), \quad (6)$$

where $w'$ indicates the weights of the model obtained after the initial supervised learning.

### Active semi-supervised fine-tuning

The model does not fully learn the main task after the initial supervised training. The active unlearning process slightly forgets the clean features and further degrades the model's performance on the main task. Due to these two reasons, fine-tuning is used to improve the model's performance.

Before semi-supervised fine-tuning, we remove labels from all samples in dataset $D'_r (D'_r = D_o - D'_c)$ to obtain dataset $D'_{ur}$. Then performing semi-supervised learning on dataset $D'_o (D'_o = D'_{ur} \cup D'_c)$. Formally, semi-supervised learning solve the following optimization problem:

$$\min_{w^*} \frac{1}{n \times \gamma} \sum_{(x,y) \in D'_c} L_1(f_{w^*}(x), y)$$
$$+ \alpha \frac{1}{n - n \times \gamma} \sum_{x \in D'_{ur}} L_2(x, w^*) \quad (7)$$
$$+ \beta \frac{1}{n} \sum_{x \in D'_o} \mathcal{R}(x, w^*)$$

where $L_2$ is semi-supervised loss function and $\mathcal{R}$ is regularization. Weight $\alpha$ and $\beta$ denotes the trade-off. In particular, we use FixMatch (Sohn et al. 2020) to perform semi-supervised learning in DLP.

It is important to note that semi-supervised learning does not re-implant the backdoor in the model. Two reasons account for this: First, the unlabeled data will be performed on strong data augmentation, ultimately invalidating the trigger; second, poisoned samples lack labels, so the model cannot learn the association between triggers and target labels.

## Experiments

### Experimental settings

*Backdoor attacks*

We consider five state-of-the-art backdoor attacks, including poison-label backdoor attacks, specifically BadNets (Gu et al. 2017), TrojanNN (Liu et al. 2018) and Blended (Chen et al. 2017), and clean-label backdoor attacks, in particular LCA (Turner et al. 2019) and SIG (Barni et al. 2019). The above attack methods are very representative, including using heuristic triggers, optimized generated triggers, improved trigger injection method, and invisible triggers by introducing adversarial perturbation and sinusoidal signal. An example of poisoned samples generated by different attacks is shown in Fig. 3.

*Backdoor defenses*

Four state-of-the-art backdoor defenses are considered as baselines, including FP (Liu et al. 2018), MCR (Zhao et al. 2020), NAD (Li et al. 2021), ABL (Li et al. 2021) and ANP (Wu and Wang 2021). They cover the mainstream backdoor removal directions: neuron-pruning based defense, mode connectivity based defense, and knowledge distillation based defense.

*Datasets*
We evaluate the performance of all defenses against attacks in two common benchmark datasets, i.e., CIFAR10 (Krizhevsky and Hinton 2009), and ImageNet subset (Deng et al. 2009). In all the experiments, WideResNet-16-1 (Zagoruyko and Komodakis 2016) serves as the base model.

*Other settings*
We adopt the default configurations described in their papers to implement the attacks and defenses mentioned above. The defense method has access to a random subset of 5% of the clean testing set if necessary. Since the LCA attack on the ImageNet subset can not be successfully reproduced following the original paper, we omit the corresponding evaluation results. For our proposed DLP, we take FixMatch (Sohn et al. 2020) as the semi-supervised method. Besides, we adopt a SGD optimizer with a momentum of 0.9 and set the batch size 128 and the learning rate 0.01 as default. Specifically, we set crucial hyper-parameters supervised learning epochs $E_1 = 10$, active unlearning epochs $E_2 = 20$, filtering rate $\gamma = 1\%$ in all experiments.

*Evaluation metrics*
As is customary in the backdoor defense literature, we compute the two metrics to evaluate the performance of the defense: 1) attack success rate(ASR): the accuracy on the poisoned dataset, and 2) clean accuracy(CA):the accuracy on the clean dataset.

The above two metrics can measure the usability-effectiveness trade-off for backdoor defense. It is crucial for the defense mechanism to have a low ASR and a high CA, indicating that it can effectively resist backdoor attacks without adversely impacting the model's performance on the main task.

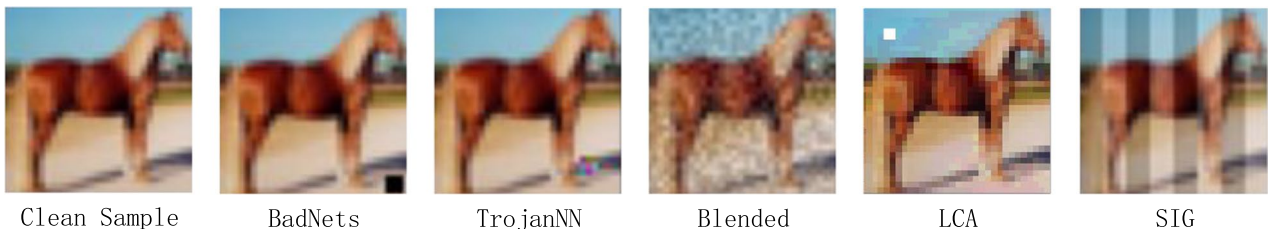

**Fig. 3** Clean sample and corresponding poisoned samples generated by different attacks

**Table 1** The defensive performance(%) of DLP against backdoor attacks on CIFAR10

| Stage | SL | | AU | | ASSFT | |
|---|---|---|---|---|---|---|
| Attack | CA | ASR | CA | ASR | CA | ASR |
| BadNets | 75.54 | 99.96 | 61.16 | 0.08 | 93.08 | 0.24 |
| TrojanNN | 77.87 | 99.78 | 65.11 | 0 | 92.19 | 0.31 |
| Blended | 71.81 | 99.99 | 45.93 | 0.02 | 91.93 | 0 |
| LCA | 74.36 | 99.86 | 70.99 | 0.04 | 92.04 | 0.21 |
| SIG | 69.04 | 100 | 52.04 | 0 | 92.98 | 0 |

**Table 2** The defensive performance(%) of DLP against backdoor attacks on ImageNet subset

| Stage | SL | | AU | | ASSFT | |
|---|---|---|---|---|---|---|
| Attack | CA | ASR | CA | ASR | CA | ASR |
| BadNets | 76.96 | 99.93 | 64.92 | 0 | 92.96 | 0.27 |
| TrojanNN | 76.49 | 100 | 62.78 | 0.04 | 93.31 | 0.21 |
| Blended | 73.92 | 100 | 45.68 | 0.07 | 93.10 | 0 |
| SIG | 76.23 | 100 | 58.32 | 0.03 | 93.08 | 0 |

## Experimental results

### Defense performance against backdoor attacks

Tables 1 and 2 summarizes the effectiveness of the three stages of DLP against backdoor attacks. After the initial supervised learning, the model achieves an average ASR of 99.92%, while the average CA is 73.72%. When active unlearning is applied, the ASR decreases significantly, whereas CAs are only slightly affected. It is evident from the performance after active unlearning that the method is very effective at eliminating backdoors. Finally, BAs can be improved to over 90% while ASRs will only get negligible improvement by semi-supervised fine-tuning. It suggests that DLP provides a good trade-off between removing the backdoor and affecting the model's performance on the main task.

As we observe, the models subjected to SIG and Blended attacks exhibit different results from the other three attacks after stage AU and stage ASSFT. Under SIG and Blended attacks, the CA of the models obtained from stage AU decreases more than the other three attacks, and the ASR of the models obtained from stage ASSFT is 0. We suspect this is because the poisoned samples produced by the above two attacks resembling mixed images, and it is hard to differentiate between backdoor features and clean features. Consequently, in stage AU, the model will unlearn both features, while in stage ASSFT, poisoned samples are more likely to be corrupted by data augmentation.

### Defense performance comparison between DLP and baselines

Tables 3 and 4 show the comparative quantitative results on the CIFAR10 and ImageNet subset, respectively. The best-performing numbers are highlighted in bold. Tables show that DLP can achieve better performance than other state-of-the-art defenses.

Before defense methods are applied, the models have both a high ASR and high CA, illustrating backdoor attacks' effectiveness. To mitigate backdoor attacks, we employ different defenses and DLP. Generally, they all work to some extent. As FP prunes neurons that are also important for the main task, higher CA must be maintained at the expense of higher ASR, so FP cannot effectively defend. In contrast, MCR, NAD, and ABL get better defense performance. However, DLP is better than them. Take the statistics on CIFAR10 as an example. BadNets's ASR can only be decreased by 95.44%, 98.01%, 95.87% with baseline methods NAD, MCR, and ABL respectively, a result worse than DLP's 99.76%.

**Table 3** Performance(%) comparison between DLP and baselines on CIFAR10

| Attack | No Attack | | BadNets | | TrojanNN | | Blended | | LCA | | SIG | |
|---|---|---|---|---|---|---|---|---|---|---|---|---|
| Defense | CA | ASR | CA | ASR | CA | ASR | CA | ASR | CA | ASR | CA | ASR |
| No defense | 93.31 | 0 | 91.31 | 100 | 88.62 | 100 | 89.89 | 100 | 87.26 | 99.46 | 88.54 | 99.88 |
| FP | 86.62 | 0 | 83.06 | 96.89 | 83.14 | 69.25 | 85.41 | 83.49 | 80.03 | 55.62 | 83.14 | 77.80 |
| MCR | 88.99 | 0 | 79.35 | 4.56 | 73.92 | 25.16 | 80.12 | 28.78 | 78.26 | 21.13 | 83.22 | 2.39 |
| NAD | 90.39 | 0 | 89.64 | 1.99 | 79.47 | 15.99 | 84.52 | 1.75 | 79.87 | 18.71 | 83.01 | 1.98 |
| ABL | 88.46 | 0 | 87.42 | 4.13 | 88.74 | 4.41 | 85.62 | 16.37 | 89.12 | 0 | 89.33 | 0.08 |
| ANP | 92.15 | 0 | 90.16 | 0.56 | 90.95 | 0.76 | 91.81 | 0.53 | 91.24 | 4.12 | 91.45 | 0.87 |
| DLP | 93.01 | 0 | 93.08 | 0.24 | 92.19 | 0.31 | 91.93 | 0 | 92.04 | 0.21 | 92.98 | 0 |

**Table 4** Performance(%) comparison between DLP and baselines on ImageNet subset

| Attack | No Attack | | BadNets | | TrojanNN | | Blended | | SIG | |
|---|---|---|---|---|---|---|---|---|---|---|
| Defense | CA | ASR | CA | ASR | CA | ASR | CA | ASR | CA | ASR |
| No defense | 93.75 | 0 | 89.99 | 100 | 90.03 | 100 | 90.64 | 100 | 90.02 | 98.85 |
| FP | 83.21 | 0 | 80.96 | 96.62 | 78.99 | 95.13 | 79.62 | 99.09 | 83.53 | 81.04 |
| MCR | 87.02 | 0 | 79.83 | 30.66 | 76.62 | 5.49 | 75.91 | 20.62 | 81.03 | 25.01 |
| NAD | 90.33 | 0 | 83.68 | 6.03 | 83.92 | 16.24 | 85.31 | 27.76 | 86.73 | 4.69 |
| ABL | 88.37 | 0 | 87.62 | 1.13 | 88.26 | 1.45 | 85.21 | 22.37 | 85.92 | 0.17 |
| ANP | 92.83 | 0 | 91.58 | 0.75 | 92.61 | 1.74 | 93.07 | 1.02 | 92.76 | 0.46 |
| DLP | 93.27 | 0 | 92.96 | 0.27 | 93.31 | 0.21 | 93.10 | 0 | 93.08 | 0 |

Furthermore, if we consider the trade-off of the defenses, DLP offers even more advantages. With DLP, CA can be maintained near the benign model level, while ASR can be reduced to almost zero.

We also compared the model's performance before and after the application of DLP under the non-attack scenario, i.e., the training dataset is not tampered, to further validate the impact of DLP on the model. For reference, a model is trained on a non-tampered training set by a standard training method. The results show no significant difference in CA between the model obtained by DLP and the model obtained by standard training. As an active defense method, DLP can train a benign model with excellent performance regardless of whether the user is vulnerable to backdoor attack.

### Further understanding of DLP
*Trade-off between ASR and CA*

DLP offers a good trade-off between ASR and CA. The model's ASR is lowered to nearly zero through active unlearning and CA is increased to the maximum through active semi-supervised fine-tuning.

Figure 4 shows the final performance of other possible similar defense settings. The ASR is hardly reduced without active unlearning (in SL-ASSFT), which means that backdoors cannot be eliminated. The CA cannot be improved without ASSFT (in SL-AU), meaning that effectiveness is sacrificed at the expense of usability to implement defense. AU and ASSFT can be arranged in reverse order and still provide an effective defense (in SL-ASSFT-AU), but the best defensive performance cannot be achieved. AU will cause the model to slightly unlearn the clean features, resulting in a further decrease in CA after the peak.

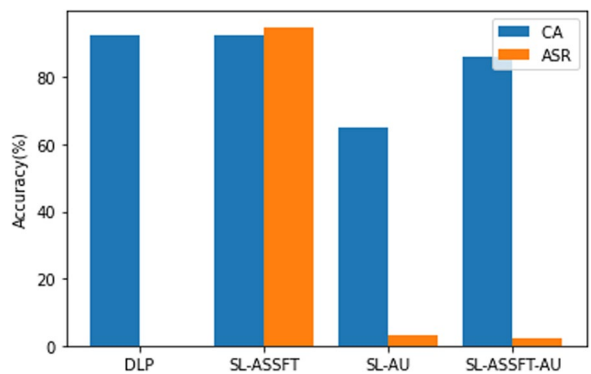

**Fig. 4** Defense performance under different defense settings

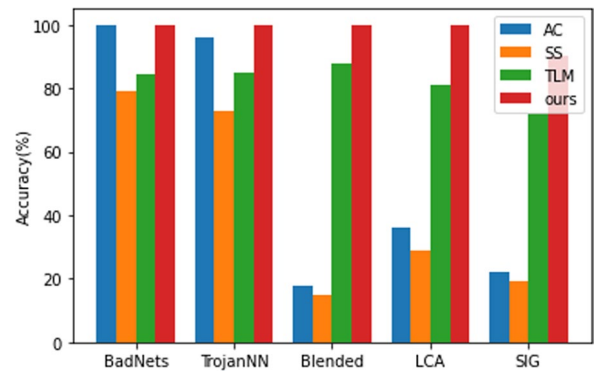

**Fig. 5** Filtering performance with different methods

By reviewing the above analysis, we can understand the rationality and effectiveness of the design of DLP.

*Effectiveness of filtering method*

In this section, we compare the effects of different methods on filtering accuracy, namely Activation Clustering (AC) (Chen et al. 2019), Spectral Signature (SS) (Tran et al. 2018), training loss-based method (TLM) and ours. AC and SS are existing defense methods for detecting poisoned samples, and TLM is the classical active learning method. In Fig. 5, we can see that our method has the highest accuracy among all attacks. We find that the accuracy of AC and SS drops a lot when detecting complex triggers since these attacks give confusing feature representations. The filtering method based on active learning is better than AC and SS, but TLM performs worse than ours. A reasonable explanation is that our method considers the results of the model's precision of all classes of the sample, which is more comprehensive compared to TLM.

**Table 5** Defense performance under different poisoning rate and filtering rate settings on CIFAR10

| Poisoning rate | 10% | | 20% | | 30% | | 40% | | 50% | |
|---|---|---|---|---|---|---|---|---|---|---|
| Filtering rate | CA | ASR | CA | ASR | CA | ASR | CA | ASR | CA | ASR |
| No defense | 91.31 | 100 | 90.22 | 100 | 89.21 | 100 | 77.92 | 99.97 | 71.49 | 100 |
| 0.10% | 87.25 | 34.26 | 86.94 | 33.08 | 84.27 | 39.37 | 81.99 | 66.9 | 81.58 | 79.23 |
| 1% | 93.08 | 0.24 | 93.01 | 2.04 | 92.88 | 3.17 | 90.54 | 3.72 | 92.06 | 3.68 |
| 5% | 91.17 | 0.24 | 93.23 | 2.06 | 92.96 | 3.05 | 91.01 | 3.59 | 90.36 | 4.99 |
| 10% | 93.88 | 0.28 | 93.59 | 1.89 | 93.61 | 3.19 | 91.32 | 2.80 | 90.81 | 4.55 |

**Table 6** Defense performance under different model architecture settings on CIFAR10

| Attack | BadNets | | TrojanNN | | Blended | | SIG | |
|---|---|---|---|---|---|---|---|---|
| Architecture | CA | ASR | CA | ASR | CA | ASR | CA | ASR |
| VGG16 | 92.43 | 0.36 | 92.47 | 0.33 | 92.26 | 0 | 92.53 | 0 |
| ResNet18 | 93.14 | 0.25 | 93.22 | 0.27 | 92.87 | 0 | 93.19 | 0 |
| InceptionV3 | 93.56 | 0.29 | 93.43 | 0.32 | 93.41 | 0 | 93.61 | 0 |
| MobileNetv2 | 93.88 | 0.19 | 93.64 | 0.25 | 93.57 | 0 | 93.92 | 0 |
| DenseNet121 | 94.82 | 0.17 | 95.02 | 0.21 | 94.96 | 0 | 95.11 | 0 |

*Effectiveness of filtering rate $\gamma = 1\%$*

Considering the current default setting of poisoning rate $\alpha = 10\%$ for mainstream backdoor attacks, we set the filtering rate $\gamma$ to 1% in the DLP.

The effect of the attacker's different ability levels (reflected in the rate of poisoning) on the defense performance is examined first. Table 5 shows that, regardless of the poisoning rate, DLP can always maintain CA over 90%. Considering removing the backdoor, we can still significantly reduce the ASR to 4% even if the poisoning rate reaches 50%. Although DLP cannot completely remove the backdoor at this time, DLP is still effective in real-world scenarios. Backdoor behavior is easier to expose as the poisoning rate increases, so backdoor attackers will not use the high poisoning rate setting in practice.

Then how the defense performance would be affected by the filtering rate is examined. The defense performance is not promising when the filtering rate is set to 0.1%, as indicated in Table 5. Only 50 samples from the CIFAR10 dataset, for instance, are being used in the training process. The experiment result is poor as a result of the few samples. On the other hand, we noticed that when the filtering rate is set to 10%, the defense performance obtained by DLP is similar to that when the filtering rate is 1%. The phenomenon indicates that the primary determinant of defensive performance is no longer the sample size.

Given the above analysis, we believe that setting $\gamma$ to 1% is reasonable.

*Generality of the method*

In this section, we experiment with other popular model architectures (VGG16 (Simonyan and Zisserman 2015), ResNet18 (He et al. 2016), InceptionV3 (Szegedy et al. 2016), MobileNetv2 (Sandler et al. 2018), DenseNet121 (Huang et al. 2017)) replacing the default WideResNet-16-1 (Zagoruyko and Komodakis 2016) while keeping other settings the same. As shown in Table 6, DLP has achieved good defense performance under different model architectures. This fully demonstrates the generality of our method.

**Conclusion**

In this paper, we identify the differential behavior of the model during backdoor learning, which led to significant differences in the prediction of the model for poisoned samples and clean samples. Based on these findings, we propose a new active defense mechanism called DLP. The DLP is based on the decoupled learning process and makes no assumption about the attack details such as poisoning rate and trigger pattern. With the DLP, one can train a benign model on the tampered dataset, preventing the model from being passively implanted the backdoor during training. Consequently, we can eliminate any further harm that may arise from the backdoored model before it occurs. Our experiments show that DLP is capable of defending against mainstream backdoor attacks and outperforms state-of-the-art defenses.

DLP is a very promising approach, but it still remains a work in progress. The main limitation of DLP is that it hardly mitigate multi-target attacks because differences in learning behavior are not reflected during its training. In future work, we will explore the commonality among backdoor attacks to develop a more comprehensive defense.

**Acknowledgements**
We would like to show our gratitude to Yige Li for sharing his codes with us. And we thank "anonymous" reviewers for their insights.

**Author contributions**
The first author completed the main work of the paper and drafted the manuscript. The second author reviewed the manuscript and revising the article critically. He also proofread the manuscript and corrected the grammar mistakes.

**Funding**
This work was supported by the National Nature Science Foundation of China under Grant No. 62272007, National Nature Science Foundation of China under Grant No. U1936119 and Major Technology Program of Hainan, China (ZDKJ2019003).

**Availability of data and materials**
The datasets used for the experiments are freely available to researchers. The links to the data have been cited as references.

## Declarations

**Competing interest**
Both authors declare that they have no competing interests.

Received: 15 September 2022   Accepted: 7 February 2023
Published online: 01 May 2023

**Publisher's Note**
Springer Nature remains neutral with regard to jurisdictional claims in published maps and institutional affiliations.